\documentclass[12pt,preprint]{aastex}

\slugcomment{}

\shorttitle{}
\shortauthors{Ling et al.}

\begin{document}

\title{Study the X-ray Dust Scattering Halo of Cyg X-1 with a Cross-correlation Method}

\author{Zhixing Ling\altaffilmark{1}, Shuang Nan Zhang\altaffilmark{1,2,3}, Jingen Xiang\altaffilmark{4}, and Shichao Tang\altaffilmark{1}}

\altaffiltext{1}{Department of Physics and Tsinghua Center for
Astrophysics, Tsinghua University, Beijing 100084, China;
lingzhixing@tsinghua.org.cn, zhangsn@tsinghua.edu.cn,
tsc02@mails.tsinghua.edu.cn}
\altaffiltext{2}{Key Laboratory of
Particle Astrophysics, Institute of High Energy Physics, Chinese
Academy of Sciences, P.O. Box 918-3, Beijing 100049, China}
\altaffiltext{3}{Physics Department, University of Alabama in
Huntsville, Huntsville, AL 35899, USA}
\altaffiltext{4}{Harvard-Smithsonian Center for Astrophysics,
Cambridge, MA 02138, USA; jxiang@cfa.harvard.edu}

\begin{abstract}

  X-ray photons scattered by the interstellar medium, carry the information of dust distribution, dust grain model,
  scattering cross section, and the distance of the source and so
  on;
  they also take longer time than the unscattered photons to reach the observer.
  Using a cross-correlation method, we study the light curves of the X-ray dust scattering halo
  of Cyg X-1, observed with the
  \textit{Chandra X-ray Observatory}. Significant time lags are found between the light curves
  of the point source and its halo.
  This time lag increases
  with the angular distance from Cyg X-1, implying a
  dust concentration
  at a distance along the line of sight of 2.0 kpc $\times$ (0.876 $\pm$
  0.002) from the Earth.
  By fitting the observed light curves of the halo at
different radii
  with simulated light curves, we obtain a width of
  $\mathit{\Delta L}=33_{-13}^{+18}$ pc of
  this dust concentration. The origin of this dust concentration is
  still not clearly known. The advantage of our method is that we need
  no assumption of scattering cross section, dust grain model, or
  dust distribution along the line of sight. Combining the derived
  dust distribution from the cross-correlation study with the
  surface brightness distribution of the halo, we conclude that the two
  commonly accepted models of dust grain size distribution need to be
  modified significantly.

\end{abstract}

\keywords{dust, extinction --- scattering --- X-rays: binaries ---
X-rays: ISM}

\section{Introduction}
The X-ray dust scattering halo was first discussed by Overbeck
(1965). Rolf (1983) first observed this phenomenon by analyzing the
data of GX339-4 with the IPC (Imaging Proportional Counter)
instrument onboard the \textit{Einstein X-ray Observatory} twenty
years later. At the same time, Catura (1983) found evidence of a
halo between 60$^{\prime \prime}$ and 600$^{\prime \prime}$ from
four low-latitude galactic X-ray sources with the HRI (High
Resolution Imager) of \textit{Einstein}. Mauche \& Gorenstein (1986)
examined four Galactic (low-latitude) sources and two extragalactic
(high-latitude) sources with the IPC of \textit{Einstein} and found
that the intensity of the halo was correlated well with the visual
extinction. They also found that the shape and fraction of the halo
derived were consistent with the common dust model, e.g., the one
established by Mathis, Rumpl \& Nordsieck (1977). After the launch
of \textit{ROSAT}, Predehl \& Schmitt (1995) analyzed the data of 25
point sources and four supernova remnants and found a strong
correlation between the visual extinction and the hydrogen column
density.

Because of the poor angular resolution of those satellites, the data
of all papers above are only contain information of the halo between
60$^{\prime \prime}$ and 1000$^{\prime \prime}$ from the center
source. After the launch of the \textit{Chandra X-ray Observatory},
it has become possible to study the halo within 10$^{\prime
\prime}$, and even 1$^{\prime
  \prime}$.  Smith, Edgar \& Shafer (2002) first reported the halo of
GX 13+1 between 50$^{\prime \prime}$ and 600$^{\prime \prime}$ with
the data of ACIS (Advanced CCD Imaging Spectrometer). Because of the
serious pileup, many observational data of ACIS cannot be used.

To avoid the effect of pile up, Yao et al. (2003) determined the
halo of Cyg X-1 as close to the point source as 1$^{\prime \prime}$
by a reconstruction method with the data of Continuous Clocking Mode
of ACIS. Xiang, Zhang \& Yao (2005) reconstructed the halo's surface
brightness of 17 bright sources and deduced the dust distribution
along the LOS (line of sight) with the data of ACIS-S array (This
array is not focused for imaging, instead it is placed on the
transmission gratings's Rowland circle. However it does not matter
here because the scattering halo is diffuse, as long as the point
source is near the nominal grating focus, and therefore properly
focused). The method used in all of papers above is to evaluate the
halo brightness distribution after removing the point spread
function (PSF) from the observed surface brightness, therefore those
results are very sensitive to the PSF. Another way to study the halo
is the effect of delay and broadening of the light curve.
Tr$\mathrm{\ddot{u}}$mper \& Sch$\mathrm{\ddot{o}}$nfelder (1973)
first proposed to use the delay and smearing property to determine
the distance of X-ray sources. Predehl et al. (2000) used the delay
property in determining the distance of Cyg X-3 with the data of
ACIS. Hu, Zhang \& Li (2004) developed a method of using the power
density spectra to determine the distances of X-ray sources. Xiang,
Lee \& Nowak (2007) used the delay property determining the distance
of 4U 1624-490. Vacua et al. (2004, 2006) found ring structures in
two GRB observations with \textit{XMM-Newton} and \textit{Swift}.
Those rings are results of the scattering of the molecular cloud
near the Sun. Shao \& Dai (2007) and Shao, Dai \& Mirabel (2008)
proposed that the X-ray afterglows of some GRBs may come from the
dust scattering near the source and they successfully modeled the
light curves of their X-ray afterglows of many GRB observations.

In this work, we use the cross-correlation method, described in
section 2, to study the light curves of the X-ray dust scattering
halo of Cyg X-1. Time lag peaks are found in the cross-correlation
curves corresponding to different observational angles from the
center of the source and different energy bands. With an assumption
of the distance of Cyg X-1 to be 2 kpc (Mirabel $\&$ Rodrigues
2003), we find that the different time lags reveal a dust
concentration at 2.0 kpc $\times$ (0.876 $\pm$
  0.002) from the observer. After modeling
the PSF with ChaRT and MARX, we remove the influence of PSF and get
the clean light curves of the source and halo simultaneously. The
time lag can also be seen in those curves directly. The dust layer's
width can be estimated by simulations. We derive a dust width of
$\mathit{\Delta L}=33_{-13}^{+18}$ pc along LOS. We show our results
in detail in section 4 and discussions in section 5.

\section{Method}

\begin{figure}
\begin{center}
  \includegraphics[angle=0,scale=.3]{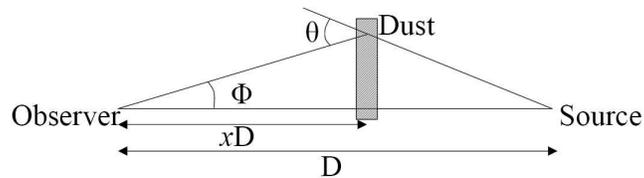}\\
  \caption{X-ray dust scattering geometry.}\label{fig:illu}
\end{center}
\end{figure}

The details of X-ray dust scattering can be found in Van de Hulst
(1957), Overbeck (1965), Tr$\mathrm{\ddot{u}}$mper \&
Sch$\mathrm{\ddot{o}}$nfelder (1973), Smith \& Dwek (1998). Here we
only discuss the single scattering because of the low halo fraction
of Cyg X-1 (Xiang, Zhang \& Yao 2005). As shown in Fig.
\ref{fig:illu}, the X-ray source is located at a distance of
$\mathit{D}$. The dimensionless number $\mathit{x}$ is the fraction
of the distance of scattering and that of the source from us. So the
lag time of scattered photons at $\mathit{x}$ can be expressed as

\begin{equation}
t_\mathrm{Delay}(\phi,x)=(\frac{x}{\cos\phi}+\sqrt{(1-x)^2+(x\tan\phi)^2}-1)\times
\frac{D}{c}.
\end{equation}
Let $\mathit{L(t)}$ denotes the luminosity of the source at the
location of the source, the observed halo intensity at different
observational angle $\phi$ is given by

\begin{equation}
I(\phi,t)=\int_{0}^{1}{{\mathrm{d}}
x}\times{D}\times\frac{L(t-t_\mathrm{Delay}(\phi,x))\times{\rho(x)}}{(1-x)^2\times4{\pi}D^2}\times\frac{\mathrm{d}\sigma(\theta)}{\mathrm{d}\Omega},
\end{equation}
here the scattering cross section
$\frac{\mathrm{d}\sigma(\theta)}{\mathrm{d}\Omega}$ depends on the
energy of the X-ray photon and the radius of the dust grain.
$\rho(x)$ is the density of the dust grain at $\mathit{x}$. From
Equation 2, the light curve of the halo is delayed and broadened
from that of the source.

We can study the delay property directly with the cross-correlation
method. The definition of cross-correlation coefficient is given by
\begin{equation}
c(\Delta t)=\frac{1}{N-\arrowvert {\Delta
t}\arrowvert}\sum_{t=0}^{N-\arrowvert {\Delta t} \arrowvert-1}  (
L_{h}(t+\Delta t) -\mu_h) (L_{s}(t)-\mu_s), \label{form:ccf}
\end{equation}
here $\mathit{L_{s}}$ and $\mathit{L_{h}}$ are the light curves of
the X-ray source and halo (at a given observational angle $\phi$) in
the same energy band. $\mu_s$ and $\mu_h$ are the average values of
$\mathit{L_{s}}$ and $\mathit{L_{h}}$ respectively.

\section{Data analysis and Simulation}

With the long exposure time and the strong variability, Cyg X-1
becomes the first target for our analysis using the
cross-correlation. This source was observed with \textit{Chandra}
HETG/ACIS-S for 47 ks on 2003 April 19 (ObsID 3814). We processed
the data using the CIAO version 3.4 and the CALDB version 3.4.1. We
first searched for point sources around Cyg X-1 using the tool
WAVDETECT of CIAO to check whether there are some other point
sources which will contaminate the light curve of the halo. The
search results show that there is no other bright point source
within the diameter of 80$^{\prime \prime}$ around the position of
Cyg X-1 during this observation. Because the halo intensity is
sensitive to the energy of photons, therefore we divide the data
into three energy bands: below 1 keV (band I), 1 keV$\sim$3 keV
(band II), and above 3 keV (band III).

\begin{figure}
\begin{center}
  \includegraphics[angle=0,scale=.3]{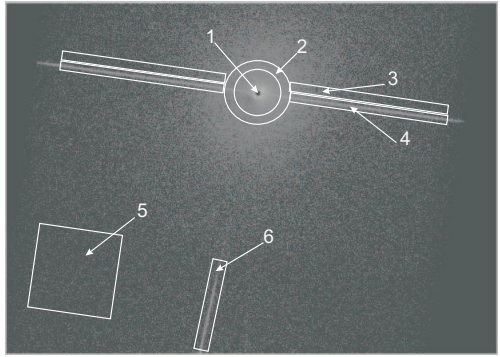}\\
  \caption{\textit{Chandra} ACIS-S image of Cyg X-1. The point
  source of this observation, shown by area 1, has no photons
  recorded because of the pileup. The annulus 2 shows the area
  for the annulus of 25$^{\prime \prime}$. The box 4 is the streak area
  and the box 3 is the background for the streak area. The box 5 is
  the background for the data of annuli. The box 6 is a grating arm.}\label{fig:streak}
  \end{center}
\end{figure}

The ACIS-S was in the 1024 pix $ \times$ 512 pix mode with a 1.74
seconds frame time during this observation. The zeroth-order data of
this observation suffered severe pileup (as shown in Fig.
\ref{fig:streak}), therefore we extracted the light curve of point
source from the zero-order streak which is caused by the charge
transfer process of the CCD (Smith, Edgar \& Shafer 2002), as the
true light curve of Cyg X-1. We extract the photons from two 200 pix
$\times$ 10 pix boxes (area 4 in Fig. \ref{fig:streak}) in the
streak area but far away from the position of Cyg X-1 to avoid the
influence of small angle scattering photons. At the same time, we
take the photons of the same area near the streak (area 3 in Fig.
\ref{fig:streak}) as the background for the streak photons. The time
bin width, which we use to extract the light curves in the whole
analysis, is 100 seconds. The background of the field of view is
taken from a box of 100 pix $\times$ 100 pix (area 5 in Fig.
\ref{fig:streak}), far away from the position of Cyg X-1 in the
field of view. After making the annuli from 5$^{\prime \prime}$ to
100$^{\prime \prime}$ with a bin step of 5$^{\prime \prime}$ (like
annulus 2 in Fig. \ref{fig:streak}) and subtracting the expected
background counts, we can extract the photons from those annuli and
obtain the light curves for each observational angle. After
extracting all light curves, we make cross-correlation curves
between the light curves of the halo and the light curve of the
source.

However, the light curves of annuli suffer from contamination of the
center point source due to the PSF of the X-ray telescope.
Therefore, we must know the exact fraction of photons in the halo
from the effect of PSF and subtract them properly. The effect of PSF
can be acquired in two different ways: 1) MARX simulation; 2) the
observations of point sources with negligible pileup and halo
contamination. The simulation is as follows: First we use Sherpa to
produce a spectrum file for ChaRT according to the parameters
acquired from the observation data we used in this work (Xiang,
Zhang \& Yao 2005). Then we use ChaRT and MARX to create the
simulated observation. The simulated observation has the same
location in ACIS with the real observation of Cyg X-1. Following the
same routine with Cyg X-1 data used before, we extract the simulated
data and obtain the photon distribution at each pixel. For method
2), the PSF fraction comes from the observation of PKS 2155-40. This
source is known as not affected by halo photons because of low
hydrogen column density  (Predehl \& Schmitt 1995). The process of
extracting the fraction of PSF is the same as that we extract from
the observation of Cyg X-1 and the data of the simulation.

\section{Results}
\begin{figure}
\begin{center}
  \includegraphics[angle=0,scale=0.5]{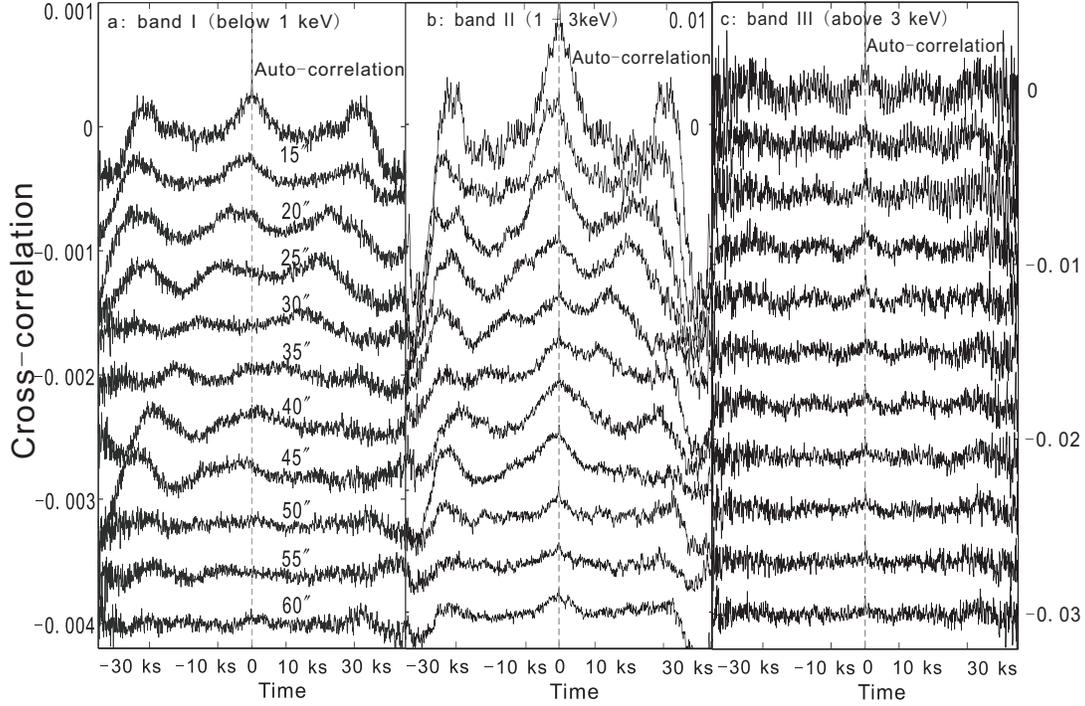}\\
  \caption{Cross-correlation curves of Cyg X-1 in band I (below 1 keV),
  band II (1$\sim$3 keV) and band III (above 3 keV). The
    top curve is the auto-correlation of the light curve of
    the source; all other curves are the cross-correlation curves from 15$^{\prime
      \prime}$ to 60$^{\prime \prime}$ with a step of 5$^{\prime
      \prime}$. For clarity, the cross-correlation coefficients have
    been lowered by a same amount (0.0004 for panel (a) and 0.004
    for panel (b) and 0.003 for panel (c)) successively for each curve. Panel (b) shows
    obvious contamination of PSF in the center of all of the cross-correlation curves, indicated by the
sharp peaks marked by the dashed line. Panel (a) suffers from less
    contamination because of the higher scattering cross section and
    narrower profile of PSF below 1 keV. Panel (c) shows more contamination than the other two band.
     }\label{fig:crs_core}
  \end{center}
\end{figure}

We compare the auto-correlation of the source light curve with the
cross-correlation between source light curve and halo light curve.
In Fig. \ref{fig:crs_core} panel (a) and (b) show two groups of
peaks in the auto-correlation curve: the peak at zero lag (Peak`0'
hereafter) with a FWHM of about 10 ks and the other two peaks about
30 ks from the center (Peak`+' for the peak of +30 ks and Peak`-'
for the -30 ks hereafter). Peak`0' is due to the variability of the
light curve at a timescale of about 10 ks, as shown in Fig.
\ref{fig:lag_direct}. The other two peaks, i.e., Peak`+' and Peak`-'
are resulted from the two big valleys separated by about 30 ks in
the light curve of the point source, as seen in Fig.
\ref{fig:lag_direct}. All these three peaks move together to the
left with increasing halo radius in both panel (a) and panel (b) of
Figure 3,  from about 1 ks to 40 ks corresponding to 5$^{\prime
\prime}$ to 50$^{\prime \prime}$ respectively. The synchronized
left-ward shift of all these three peaks in the cross-correlation
curves demonstrate the reliability of the cross-correlation method
for studying the X-ray scattering halo. Because Peak`0' is more
pronounced than Peak`+' and Peak`-', we only carry out quantitative
analysis of the lag time for Peak`0' in the following.

Panel (b) shows obvious contamination of PSF in the center of all of
the cross-correlation curves, indicated by the sharp peaks marked by
the dashed line. Panel (a) suffers from less contamination than
panel (b). Panel (c) shows more contamination than the other two
band. This phenomenon confirms that the scattering of low energy
photons are more efficient than the scattering of high energy
photons. It is not unexpected that the lag peaks disappear in band
III because of its low count rate and the smaller scattering cross
section at high energy could decrease the intensity of the halo.

\begin{figure}
\begin{center}
  \includegraphics[angle=0,scale=.5]{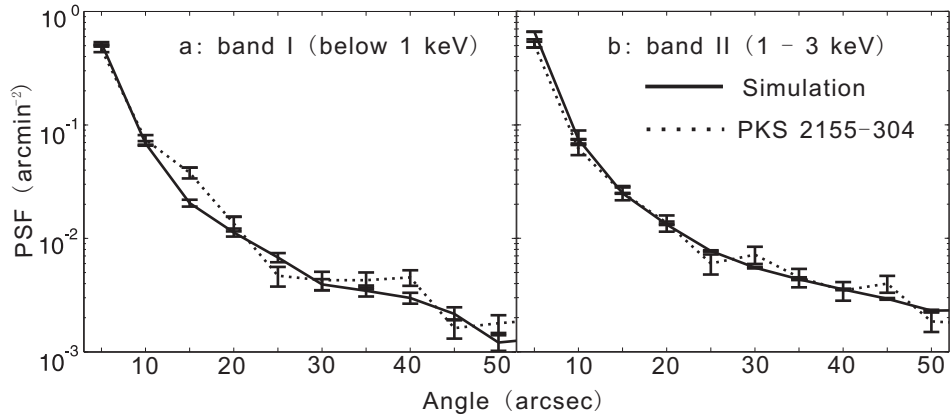}\\
  \caption{The distribution of PSF in band I and band II. The dotted
  line shows the observed result of PKS 2155-304. The solid line
  shows the simulated result. This figure shows that the PSF of low
  energy photons
  is narrower than the high energy photons.}\label{fig:psf}
  \end{center}
\end{figure}

The observed and simulated radial profiles of PSF are shown in Fig.
\ref{fig:psf}. The dotted line shows the result of observation of
PKS 2155-304 (ObsID 3807). The solid line shows the simulated
result. We found that there is no obvious difference between those
two results. However, as the locations of the sources of different
observations are not always at the same position of the ACIS during
the observations of PKS 2155-304, in the following we only use PSF
with simulated data, since we can set the positions in simulations.

\begin{figure}
\begin{center}
  \includegraphics[angle=0,scale=.7]{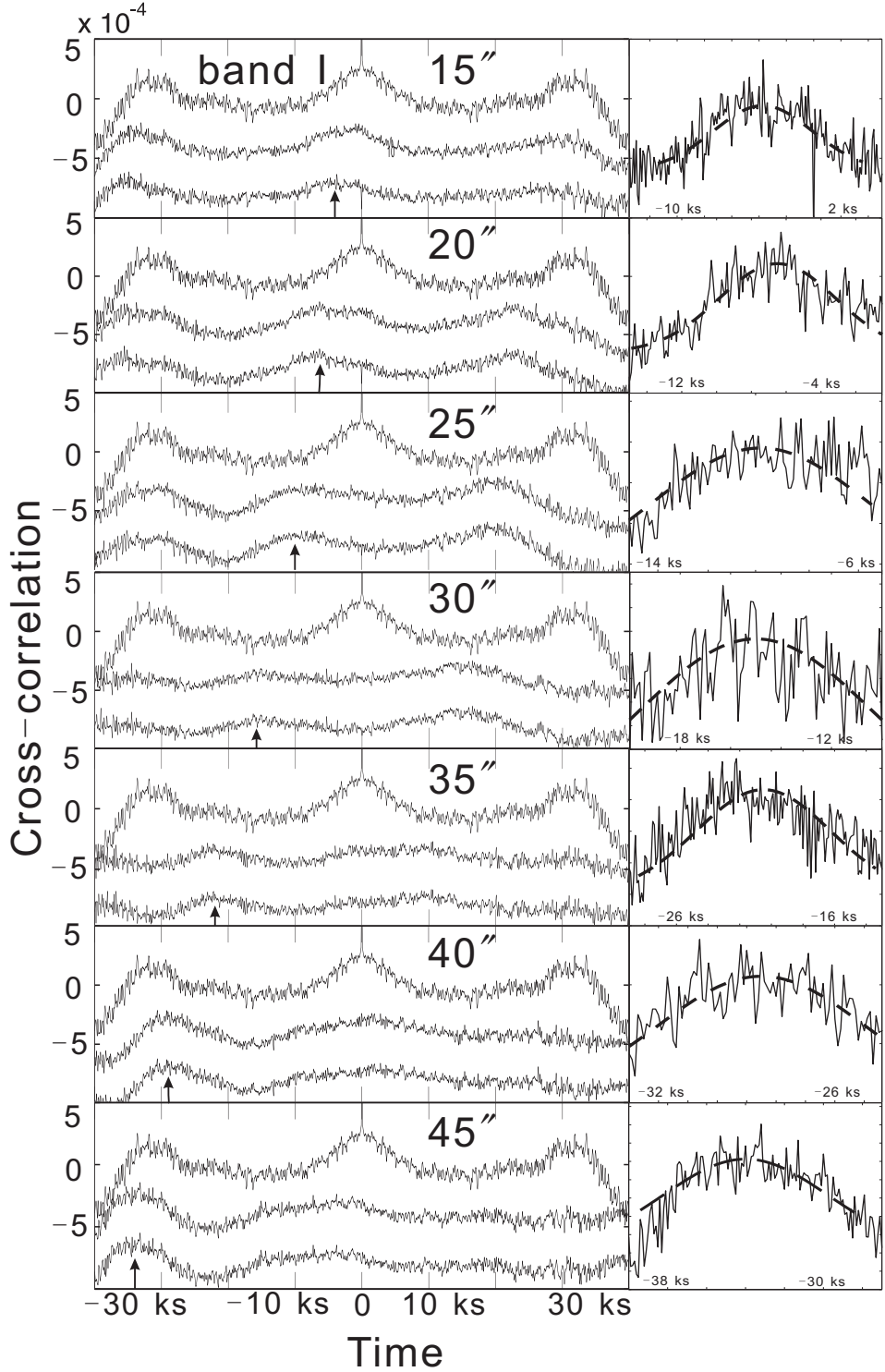}\\
  \caption{Cross-correlation curves of the halo from 15$^{\prime
      \prime}$ to 45$^{\prime \prime}$ of band I. The top curve in
    each panel is the auto-correlation of the light curve of the
    source, the middle curve is the cross-correlation of the light curve
    of the halo and that of the source, the bottom curve shows the
    same with the middle curve after removing the contamination of PSF
    effect. The lower two curves have been lowered by 0.0004 for
    clarity. The arrows in each panel indicate the lag time at
    each annulus. The fitting result of the peaks are
    shown in the right of each sub-figure, with a simple Gaussian function.}\label{fig:lag_1}
  \end{center}
\end{figure}

\begin{figure}
\begin{center}
  \includegraphics[angle=0,scale=.7]{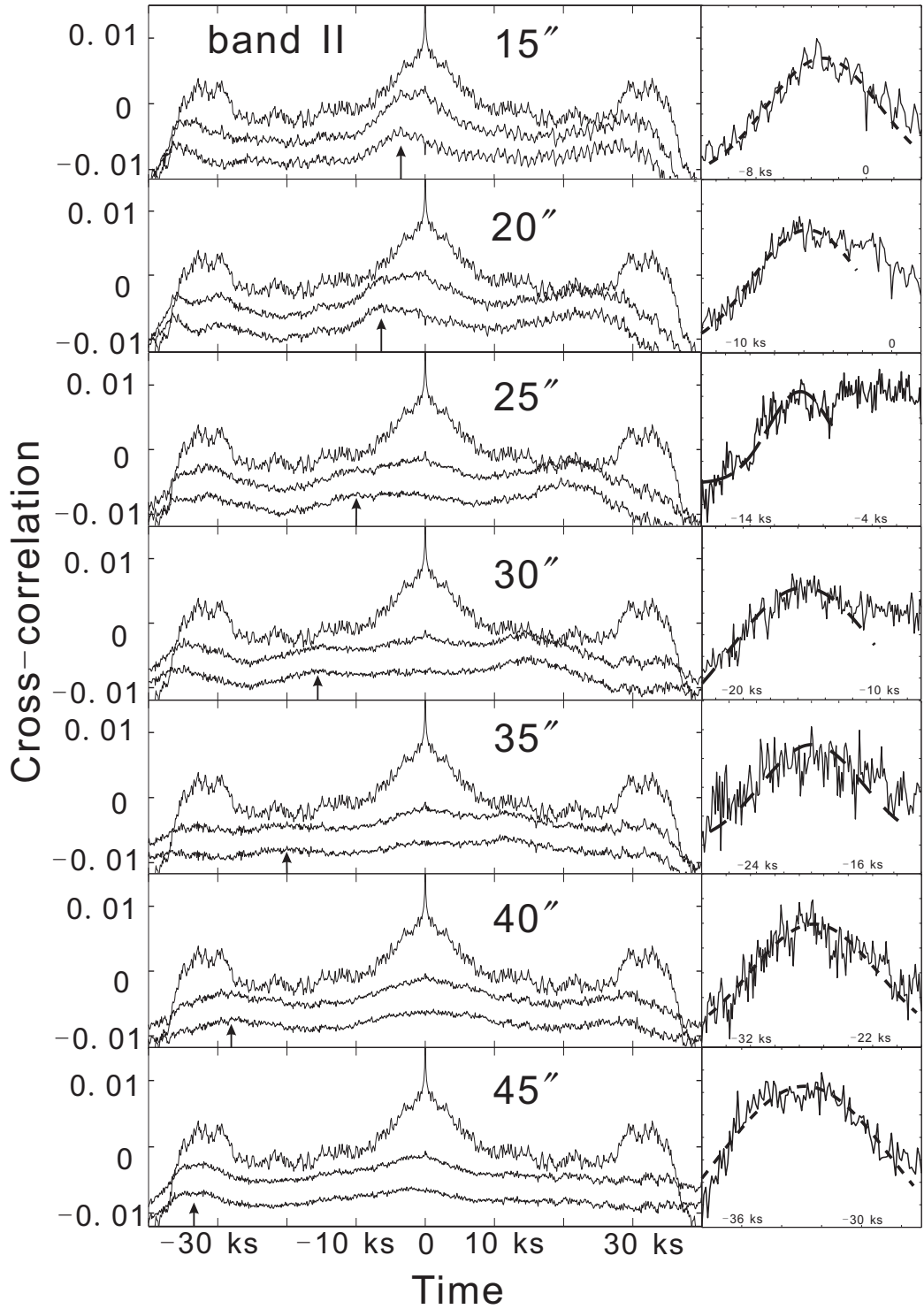}\\
  \caption{The same as Fig. \ref{fig:lag_1} for band II (1$\sim$3 keV).
  In this figure, the lower two curves have been lowered by 0.004 for
  clarity. The lag peaks of 20$^{\prime \prime}$, 25$^{\prime \prime}$
  and 30$^{\prime \prime}$ are fitted only with the left data of the peak.
  The asymmetry of those peak may come from the unremoved contamination of PSF.} \label{fig:lag_2}
  \end{center}
\end{figure}

The total counts of the source are estimated with the number of
streak counts times the exposure factor, which equals to the frame
time divided by the total transfer time of the streak area we used
($n_\mathrm{source}=n_\mathrm{streak}\times 1.74~  \mathrm{s} / (40~
\mu  \mathrm{s} /  \mathrm{pix}\times 400~  \mathrm{pix})$). After
the subtraction of the PSF, the total intensity of the halo from
10$^{\prime \prime}$ to 100$^{\prime \prime}$ is about
6.37$\pm$0.04$\%$, which is consistent with Xiang, Zhang \& Yao
(2005). With the PSF contamination removed, we can then derive the
real time lag of each annulus. Fig. \ref{fig:lag_1} and Fig.
\ref{fig:lag_2} show the cross-correlation curves of 15$^{\prime
\prime}$ to 45$^{\prime \prime}$ in band I and band II. The top
curve in each panel is the auto-correlation of the light curve of
the source, the middle curve is the cross-correlation of the light
curve of the halo and that of the source, the bottom curve shows the
same with the middle curve after removing the contamination of PSF
effect. The lower two curves have been lowered for clarity. The
arrows in each panel indicate the lag time at each halo radius. The
fitting result of those peaks are shown in the right of each
sub-figure, with a simple Gaussian function. The lag peaks of
20$^{\prime \prime}$, 25$^{\prime \prime}$ and 30$^{\prime \prime}$
in Fig. \ref{fig:lag_2} are fitted only with the left data of the
peak. The asymmetry of those peaks may come from the unremoved
contamination of PSF. In Fig. \ref{fig:lag_1}, the cross-correlation
curves also show the lag peaks clearly even before the elimination
of PSF (the middle curve of each sub-figure in Fig.
\ref{fig:lag_1}).

As shown in Fig. \ref{fig:lag_1} and Fig. \ref{fig:lag_2}, Peak`+'
in the cross-correlation curves moves to the left, and Peak`-' moves
out of the cross-correlation curves. The intervals of those peaks in
the bottom curve of each panel are always the same as in the
auto-correlation, about 30 ks, as expected by applying the
cross-correlation method.

\begin{figure}
  \includegraphics[angle=0,scale=.6]{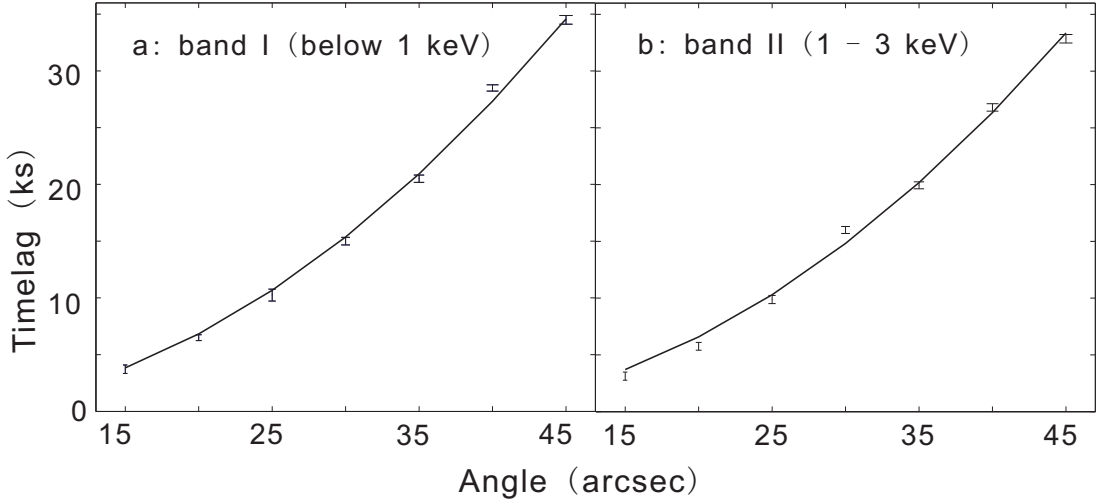}\\
  \caption{The lag time vs halo radius. Panel (a) shows the
    result of band I (below 1 keV) and panel (b) shows band II (1$\sim$3 keV). The best fit result
    is $\mathit{x}$ = 0.876$\pm$ 0.002 for band I and $\mathit{x}$ =
    0.872$\pm$ 0.002 for band II. The solid lines show the best fit
    results. Because of its less contamination of PSF and better
    fitting, the result of band I is used in this work.
    }\label{fig:location}
\end{figure}

\begin{figure}
  \includegraphics[angle=0,scale=.6]{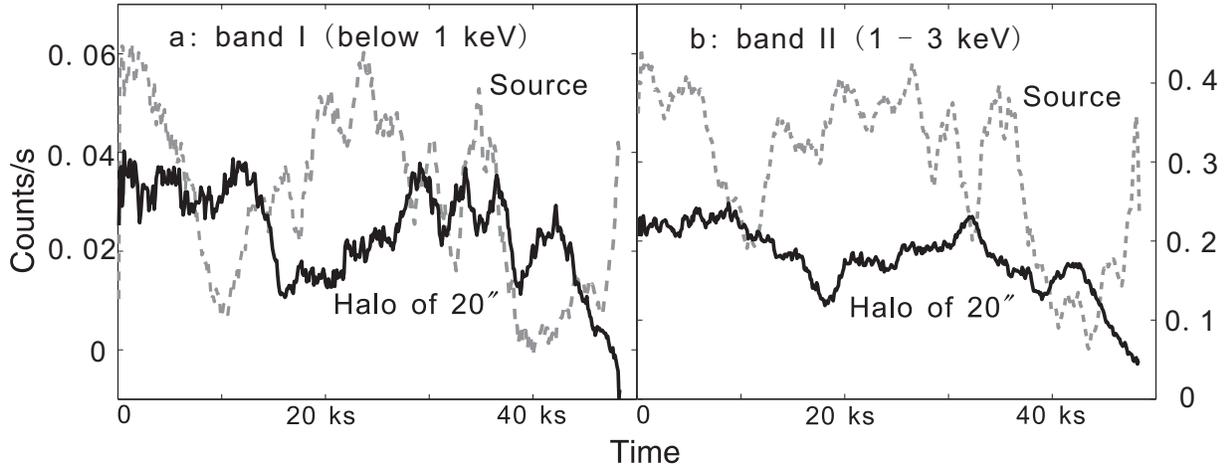}\\
  \caption{Light curve of the source and its halo of 20$^{\prime
      \prime}$ in band I and band II. The gray dashed line shows the light curve of the
    source. The black solid line shows the light curve of the halo of
    20$^{\prime \prime}$. Here the influence of PSF have been removed
    from the light curve of the halo. Both the light curves have been smoothed with a window of 2
    ks, in order to suppress the counting fluctuations. A lag of around 8 ks is seen clearly.}\label{fig:lag_direct}
\end{figure}

By fitting the Peak`0' in each cross-correlation curves, we can get
the time lags at each halo radius. The results are shown in Fig.
\ref{fig:location}. We use a simple dust wall model to fit the
different time lag, with an assumption of 2 kpc  distance from Cyg
X-1 to us (Mirabel $\&$ Rodrigues 2003). Panel (a) shows the result
of band I (below 1 keV) and panel (b) shows band II (1$\sim$3 keV).
The best fit result is $\mathit{x}$ = 0.876$\pm$ 0.002 for band I
and $\mathit{x}$ = 0.872$\pm$ 0.002 for band II. The solid lines
show the best fit results. Because of its less contamination of PSF
and better fitting, the result of band I is used in this work. The
result shows that the dust exists at $\mathit{x}$ = 0.876 $\pm$
0.002, i.e., 1.752 kpc away from us.

To illustrate the time lag revealed by the cross-correlation method,
we plot the light curves of the source and halo together. As shown
in Fig. \ref{fig:lag_direct}, the light curves of the source and the
halo at 20$^{\prime \prime}$ in band I and band II both reveal an
obvious lag of about 8 ks, just as the result from Fig.
\ref{fig:location}. Both the light curves in Fig.
\ref{fig:lag_direct} have been smoothed with a window of 2 ks, in
order to suppress the counting fluctuations.

\begin{figure}
\begin{center}
  \includegraphics[angle=0,scale=0.9]{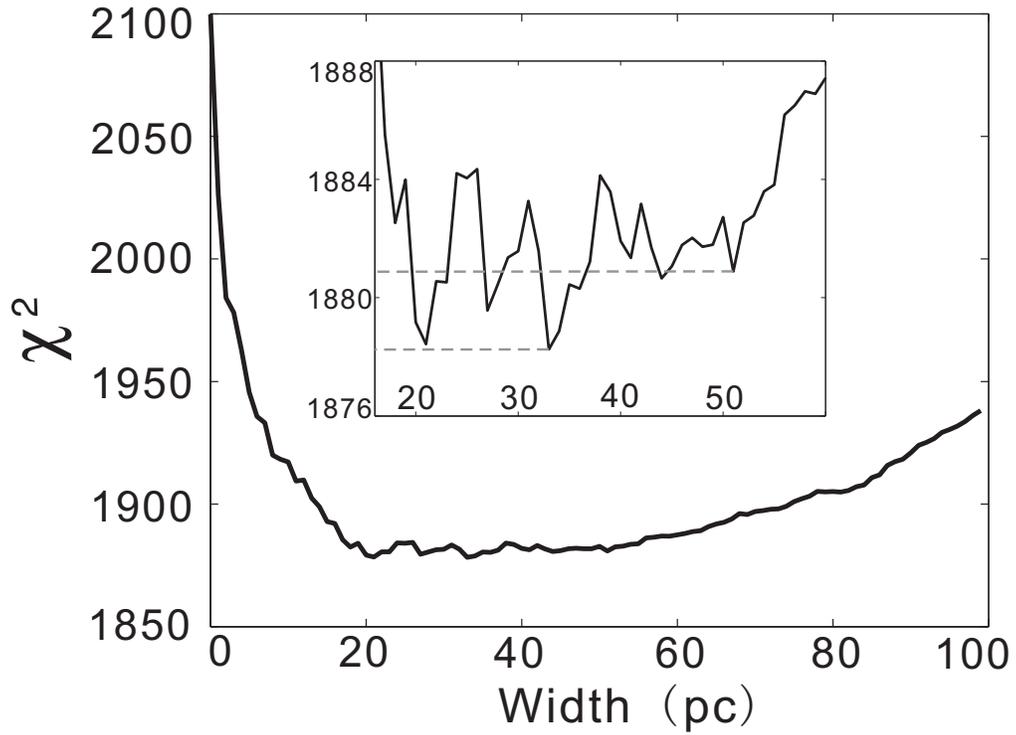}\\
  \caption{The $\chi^2$ distribution vs the width of dust layer $\mathit{\Delta L}$.
  We use the light curves of band II here because of its high count rates. The minimum
    $\chi^2$ = 1878 with 1664 degrees of freedom. The 90$\%$ confident range
    is [20, 51] pc, with the $\Delta\chi^2$ = 2.71 (Avni 1976).}\label{fig:chi2}
  \end{center}
\end{figure}

\begin{figure}
\begin{center}
  \includegraphics[angle=0,scale=0.9]{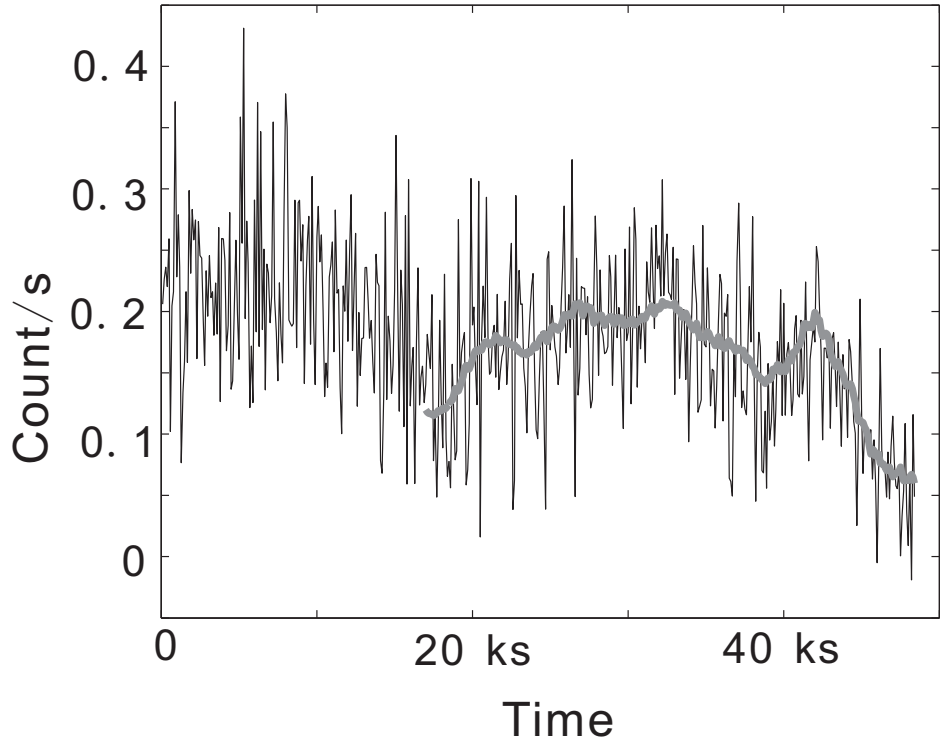}\\
  \caption{The observed light curve of the halo of 20$^{\prime \prime}$. The grey line shows
    the simulated result for $\mathit{\Delta L}$ = 33 pc and $\mathit{x}$ = 0.876.
    Clearly the simulated light curve resembles the observed light
curve quite well. The simulated light curves at each time bin number
$\mathit{i}$ is depending on the observational light curve before
the time bin number $\mathit{i}$ with a length of the response
function because of the convolution process. However, since the
light curves are not available before the start of this observation,
the initial part of the simulated data of about 16 ks is
omitted.}\label{fig:light_simu}
  \end{center}
\end{figure}

The width of the dust wall at $\mathit{x}$ = 0.876 can also be
estimated from the broadening property of the light curves of the
halo. We use a simple model of a dust layer extended from
$\mathit{x}$ = 0.876 to both sides along the LOS to simulate the
light curves of the halo. With Equation 2, we could obtain the ideal
response curve of a delta function with the parameter of the dust
layer width of $\mathit{\Delta L}$. After convolving the light curve
of the source with the ideal response function, we could produce
light curves of the halo at each angle. We define the $\chi^2$ by
\begin{equation}
\chi^2=\sum_{i}(\frac{n_\mathrm{i, simu}-n_\mathrm{i,
obser}}{\sigma_\mathrm{i, obser}})^{2},
\end{equation}
here $\mathit{i}$ is the time bin number including the light curves
at each angle. $n_\mathrm{i, simu}$ and $n_\mathrm{i, obser}$ denote
the simulated and observed count rates at each time bin number
$\mathit{i}$. $\sigma_\mathrm{i, obser}$ denotes the error of
observed data at time $\mathit{i}$. We use the observed light curves
of band II here because of its high count rates. By comparing the
simulated curves with the observed curves, we get a range of the
parameter $\mathit{\Delta L}$. As shown in Fig. \ref{fig:chi2}, the
minimum $\chi^2$ = 1878 with 1664 degrees of freedom. The 90$\%$
confident range is [20, 51] pc, with $\Delta\chi^2$ = 2.71 (Avni
1976). We show the simulated light curve of 20$^{\prime \prime}$ and
the observed light curve of 20$^{\prime \prime}$ in Fig.
\ref{fig:light_simu} with $\mathit{x}$ = 0.876 and $\mathit{\Delta
L} = 33$ pc. Clearly the simulated light curve resembles the
observed light curve quite well. The simulated light curves at each
time bin number $\mathit{i}$ is depending on the observational light
curve before the time bin number $\mathit{i}$ with a length of the
response function because of the convolution process. However, since
the light curves are not available before the start of this
observation, the initial part of the simulated data of about 16 ks
(this value means that the scattering happens at $\mathit{x}$ =
0.95, which corresponds to the largest distance the dust layer can
reach) is omitted in Fig. \ref{fig:light_simu}.

\begin{figure}
\begin{center}
\includegraphics[angle=0,scale=0.6]{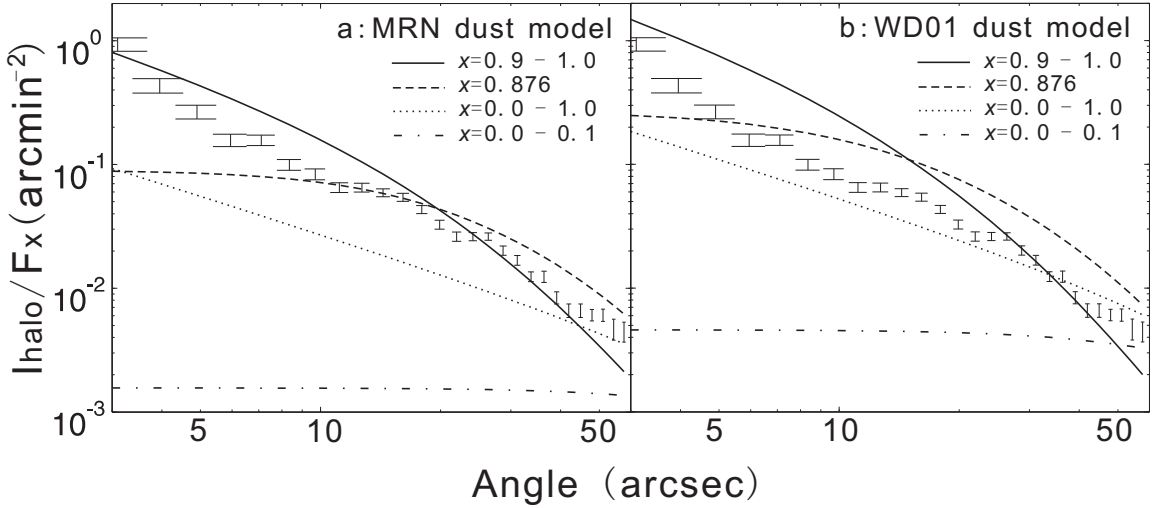}\\
\caption{Halo profile of Cyg X-1 in the energy band of 1.0 keV to
5.0 keV (Xiang, Zhang $\&$ Yao 2005). Here we use the MRN dust model
(panel a) and WD01 dust model (panel b) with a hydrogen column
density of $5.0\times 10^{21}/\mathrm{cm^2}$. The dashed line is the
result of dust in 0.876. The solid line is the result of dust smooth
distributed between $\mathit{x}$ = 0.9 to 1.0. The dotted line is
the result of dust smooth distributed between $\mathit{x}$ = 0.0 to
1.0. The dashed dotted line is the result of dust smooth distributed
between $\mathit{x}$ = 0 to 0.1. From those curves, we can infer
that the fitting of halo surface brightness with MRN or WD01 dust
models would lead to a dust concentration near the
source.}\label{fig:notfit}
\end{center}
\end{figure}

However, Xiang, Zhang $\&$ Yao (2005) fitted the halo surface
brightness of 17 sources with two dust grain models and claimed that
most of the dusts along the LOS of Cyg X-1 are located at
$\mathit{x}$ larger than 0.9. To explain this difference, we plot
the halo surface brightness of Cyg X-1 again. As shown in Fig.
\ref{fig:notfit}, we find that the halo surface brightness cannot be
fit with the dust distribution in which the dusts are located at
$\mathit{x}$ = 0.876 or dust uniformly distributed between
$\mathit{x}$ = 0 and $\mathit{x}$ = 1. The halo surface brightness
of Cyg X-1 of Xiang, Zhang $\&$ Yao (2005) would lead to a dust
concentration near the source with the MRN (Mathis, Rumpl \&
Nordsiech 1977) dust grain model or the WD01 (Weingartner $\&$
Draine 2001) model. Witt, Smith, $\&$ Dwek (2001) fitted the halo
surface brightness by the MRN dust grain model and a smoothed dust
distribution. They found that the halo of Nva Cygni 1992 cannot be
fitted by the MRN model unless the model is modified with the
maximum radius of dust grain extending to 2 $\mu$m. Valencic and
Smith (2008) also argued that the radius of dust grain of MRN or
WD01 model are not realistic. They tested the MRN and WD01 dust
grain models using the UV extinction and X-ray halos along the LOS
toward X Per. Both models provided reasonable fits to the UV
extinction and X-ray halos but cannot do so while respecting the
elemental abundance constraints. Furthermore, the abundances and
$N_H$ required to reproduce the observations in these two regimes
are not consistent with each other, reflecting the fact that X-ray
regime constraints were not taken into account when the grain size
distributions were constructed, and thus the models are incomplete.
The MRN model has long been known to be unphysical and WD01 model
was simply never designed with the X-ray regime, as pointed out by
Valencic and Smith (2008). All results cast doubt on those dust
models. We therefore conclude that the two commonly accepted models
of dust grain size distribution must be modified significantly
before they are used in the X-ray regime.

\section{Summary and Discussion}

We applied the cross-correlation method to the light curves of Cyg
X-1 and found the photons of the halo significantly lag from that of
the source itself. This time lag implies a dust concentration at
$\mathit{x}$ = 0.876 $\pm$ 0.002 with the source distance of 2.0
kpc. But the lag disappears when the energy of photons exceeds 3 keV
in this analysis: both the low count rate and the smaller scattering
cross section could decrease the intensity of halo. By the
broadening property of light curves, we derived a dust width of
$\mathit{\Delta L} = 33_{-13}^{+18}$ pc.

The cross-correlation method can avoid assumptions of photon energy,
dust grain radius, dust distribution, scattering cross section and
so on. Therefore the time lag derived by this method only rests on
pure geometry, but not connected at all with the parameters of dust
grain. Consequently, our results can be used to determine the
parameter of the dust grain models in the future, when combined with
the spatial distribution of the X-ray dust scattering halo;
currently no dust grain model can describe simultaneously the time
lag and spatial distribution of X-ray dust scattering halo. Besides,
if the position of the scattering dust cloud is known from other
methods, we can also derive the distances of the sources with this
cross-correlation method.

The origin of this dust layer is still unknown. There may exists a
molecular cloud. The width of $\mathit{\Delta L} = 33_{-13}^{+18}$
pc is also consistent with a typical molecular cloud. We searched
for all of the known molecular clouds but failed to find a
counterpart for this dust layer. Another possibility is that there
is a super-bubble around Cyg X-1, and $\mathit{x}$ = 0.876 $\pm$
0.002 is just the edge along the LOS. This hypothesis is consistent
with Gallo et al. (2003) who found a low density region of about 5
pc around Cyg X-1. We also did not find longer time lags
corresponding to the region of about 5 pc to the source in
cross-correlation curves or light curves, indicating there is very
little dust between $\mathit{x}$ = 0.876 and the center source along
the LOS. We also cannot constrain well the dust distribution between
the inferred dust wall and the observer. This is because that the
scattered photons from any dust in this region are not significant
for the halo at smaller angular  distances. The only qualitative
statement about the dust between the wall and the observer is that
the dust density outside the wall must be much lower than in the
wall; otherwise shorter lags should have been detected in the halo's
light curve. Finally, we did not intend to derive the dust and
neutral hydrogen column density, because the dust grain size
distribution and scattering cross section are not well understood
yet.

\acknowledgments

We thank Jian Hu, Yuan Liu, Yue Wu, and Li Shao for useful
discussions and Randall K. Smith for providing the model codes. The
anonymous referee is thanked for kind suggestions which helped to
clarify several points. This study is supported in part by the
Ministry of Education of China, Directional Research Project of the
Chinese Academy of Sciences under project No. KJCX2-YW-T03 and by
the National Natural Science Foundation of China under project No.
10521001, 10733010 and 10725313.

\end{document}